\documentclass[aps]{revtex4}  
\usepackage{graphicx}  
\usepackage{dcolumn}   
\usepackage{bm}        
\usepackage{amssymb}   
\def\MET{{\mbox{$E\kern-0.57em\raise0.19ex\hbox{/}_{T}$}}}
\def\met{{\mbox{$E\kern-0.57em\raise0.19ex\hbox{/}_{T}$}}}
\def\DZ{D\O\ }
\def\DZero{D\O\ }
\def\Dzero{D\O\ }

\def\ifb{~fb$^{-1}$}
\def\pp{$p\bar{p}$}

\def\ttbar{$t\bar{t}$}
\def\WH{$WH\rightarrow \ell\nu b\bar{b}$}
\def\lmet{$WH\rightarrow \ell\kern-0.45em\raise0.19ex\hbox{/} \nu b\bar{b}$}
\def\ZH{$ZH\rightarrow \nu\bar{\nu} b\bar{b}$}
\def\ZHll{$ZH\rightarrow \ell^+ \ell^- b\bar{b}$}

\def\pwh{$p\bar{p}\rightarrow WH \rightarrow \ell \nu b\bar{b}$}
\def\pzh{$p\bar{p}\rightarrow ZH \rightarrow  \nu\bar{\nu} b\bar{b} / \ell^+\ell^- b\bar{b}$}

\def\pwww{$p\bar{p}\rightarrow WH \rightarrow WW^{+} W^{-}$}
\def\www{$WH \rightarrow WW^{+} W^{-}$}
\def\phww{$p\bar{p}\rightarrow H \rightarrow W^{+} W^{-}$}
\def\hww{$H\rightarrow W^+ W^-$}
\def\hbb{$H\rightarrow b\bar{b}$}

\def\tevE{$\sqrt{s}=1.96$~TeV}

\begin{document}
\rightline{FERMILAB-PUB-08-069-E}
\rightline{CDF Note 9290}
\rightline{\DZ Note 5645}
\vskip0.5in

\title{Combined CDF and \DZ Upper Limits on Standard Model Higgs-Boson Production with up to 2.4 fb$^{-1}$ of data\\[2.5cm]}

\author{
The TEVNPH Working Group\footnote{The Tevatron
New-Phenomena and Higgs working group can be contacted at
TEVNPHWG@fnal.gov. More information can be found at http://tevnphwg.fnal.gov/.}
 }
\affiliation{\vskip0.3cm for the CDF and \DZ Collaborations\\
\vskip0.2cm 
\today} 
\begin{abstract}
\vskip0.3in
We combine results from CDF and D\O\ searches for a standard model Higgs
boson ($H$) in  \pp~collisions at the Fermilab Tevatron at
$\sqrt{s}=1.96$~TeV.
 With 1.0-2.4\ifb~ of data collected at CDF, and
1.1-2.3\ifb~at D\O, the 95\% C.L. upper limits 
on  Higgs boson production 
are a factor
of 3.7~(1.1) higher than the SM cross section for a Higgs boson mass of
$m_{H}=$115~(160)~GeV/c$^2$. 
Based on simulation, the median expected upper limit should be
3.3~(1.6). 
Standard Model branching ratios, calculated as functions of
the Higgs boson mass, are assumed.
Compared to the previous Higgs Tevatron combination, more data 
and new channels ($H \rightarrow \gamma \gamma$, and $H \rightarrow
\tau \tau$ produced in several modes) have been added.
Existing channels,
such as both experiments' $ZH\rightarrow \nu{\bar{\nu}} b{\bar{b}}$ channels, have
been reanalyzed to gain sensitivity.
These results extend significantly the
individual limits of each experiment.\\[2cm]
 {\hspace*{5.5cm}\em Preliminary Results}
\end{abstract}

\maketitle

\newpage
\section{Introduction} 
The search for a  mechanism for electroweak symmetry breaking, and in 
particular
for a standard model (SM) Higgs boson boson has been a major goal
of High Energy Physics for many years, and
 is  a central
part of Fermilab's Tevatron program. Both CDF and \Dzero 
have recently 
reported new searches for the SM Higgs boson that combined different production and decay modes~\cite{CDFhiggs,DZhiggs}
that allow to gain sensitivity compared to the previous Tevatron combination presented in December 
2007~\cite{tev-comb-2007}.
In this note, we combine the most recent 
results of all such searches 
in \pp~collisions at~\tevE. The searches for a SM Higgs
boson produced in association with vector bosons 
(\pwh, \pzh~or \pwww) or
through gluon-gluon fusion (\phww) or vector boson fusion (VBF), in data
corresponding to integrated luminosities ranging from 1.0-2.4\ifb~at
CDF and 1.1-2.3\ifb~at D\O . 
In this combination we add for the first time searches for Higgs bosons
decaying to two photons or two tau leptons.

To simplify their combination, 
the searches are separated into twenty nine
mutually exclusive 
final states (thirteen for CDF, sixteen for D\O , see
Table~\ref{tab:cdfacc} and ~\ref{tab:dzacc}) 
referred to as ``analyses'' in this note. 
Selection procedures for each
analysis are detailed in Refs.~\cite{cdfWH}-\cite{dzHWW}, and 
are briefly described below.

Compared to the results that we presented at the Moriond '08                                                                    
conference~\cite{talk-moriond-qcd}, we adopt here a simpler approach in the                                                
treatment of the shape systematics uncertainties when deriving the final                                              
limits, treating them as Gaussian errors,
 as it is done for the other systematics uncertainties,
 instead of truncating them at                                                
$\pm 1 \sigma$.  This treatment is now consistent between the two limit                                                   
setting methodologies described below.  This change has essentially no                                                
effect on the expected limits and only affects the observed limits at low                                             
masses.

\section{Acceptance, Backgrounds and Luminosity}  

Event selections are similar for the corresponding CDF and D\O\ analyses.
For the case of \WH, an isolated lepton (electron or muon) and 
 two jets are required, with one or more $b$-tagged jet, i.e. identified 
as originating from a $b$-quark.  
Selected events must also display a significant
imbalance 
in transverse momentum
(referred to as missing transverse energy or \met).  Events with more than one
isolated lepton are vetoed.  
For the D\O\ \WH\ analyses,
two non-overlapping $b$-tagged samples are
defined, one being a single ``tight'' $b$-tag (ST) sample, and the other a 
double ``loose'' $b$-tag (DT) sample. The tight and loose $b$-tagging criteria
are defined with respect to the mis-identification 
rate that the $b$-tagging algorithm yields for light quark or gluon jets 
(``mistag rate'') typically $\le 
0.5\%$ or $\ge 1.5\%$, respectively.  For the CDF \WH\ analyses, an analysis based
on a sample with
two tight b-tags (TDT) is combined with an analysis based on a non-overlapping 
sample requiring 
one tight b-tag and one loose
b-tag (LDT). For this combination additional channels have been added: a single $b$-tag
channel (STC) with a dedicated neural network rejection of $c$-quark tagging , and three
additional channels with similar $b$-tagging requirement, but with electrons detected
in more forward directions than in the standard channels.
In the \WH\ analyses, both CDF and D\O\ use
neural-network (NN) discriminants as the final variables
for  setting limits.  The networks 
are optimized to discriminate signal from background
at each value of the Higgs boson mass
 (the ``test mass'') under study.

For the \ZH\ analyses, the selection is
similar to the $WH$ selection, except all events with isolated leptons are vetoed and
stronger multijet background suppression techniques are applied. 
The CDF  analysis uses two non-overlapping samples of 
events (TDT and LDT as for $WH$) while D\O\
uses a sample of events having one tight $b$-tag jet and one loose $b$-tag jet.
As there is a sizable fraction of \WH\ signal in which the lepton is undetected,
that is selected in the \ZH\ samples,  
the D\O\ analysis includes this fraction as
a separate search, referred to as \lmet.  CDF includes it as 
this fraction is included as
part of the acceptance of the \ZH~search. 
Compared to the previous combination, CDF is now also using a
neural-network discriminant as final variable, while D\O\ has doubled
the analyzed statistics and uses now boosted decision trees instead of
neural networks as advanced analysis technique.

The \ZHll\ analyses require two isolated leptons and
at least two jets. They   use non-overlapping samples of
events with one tight $b$-tag and two loose $b$-tags. 
For the D\O\ analysis a neural-network discriminant is the 
final variable for setting  limits, while  CDF uses the output
of a 2-dimensional neural-network. These analyses have not been updated compared 
to the previous combination.

 For the \hww~analyses, a large \met~and two opposite-signed, isolated leptons
(any combination of electrons or muons) are selected, defining three final states
($e^+e^-$, $e^\pm \mu^\mp$, and $\mu^+\mu^-$) for D\O .
CDF separates the \hww\ events in two non-overlapping samples, 
one having a low signal/bacgkround (S/B)
ratio, the other having a higher one.  The presence of
neutrinos in the final state prevents  reconstruction of the
Higgs boson mass. The final discriminants are neural-network outputs 
including likelihoods constructed from matrix-element probabilities as input
to the neural network, for both CDF and D\O . All analyses in this channel have
been updated with more data and analysis improvements.

The CDF experiment contributes also a new analysis searching for Higgs bosons decaying
to a tau lepton pair, in three separate production channel with 2 fb$^{-1}$ of data:
direct $p \bar{p} \rightarrow H$ production, associated $WH$ or $ZH$ production,
or vector boson production with $H$ and forward jets in the final state.
In  this analysis, the final variable for setting  limits is
a combination of several neural-network discriminants.

The D\O\ experiment
contributes three \www~analyses, where the associated $W$ boson and
the $W$ boson from the Higgs boson decay which has the same charge are required 
to decay leptonically,
thereby defining three like-sign dilepton final states 
($e^\pm e^\pm$, $e^\pm \mu^\pm$, and $\mu^{\pm}\mu^{\pm}$)
containing all decays of the third
$W$ boson. In  this analysis, which has not been updated for this combination,
the final variable is a likelihood discriminant formed from several
topological variables. 
D\O\ also contributes a new analysis searching for direct Higgs boson production
decaying to a photon pair in 2.3 fb$^{-1}$ of data.
In  this analysis, the final variable is the invariant mass of the two photons system.

All Higgs boson signals are simulated using \textsc{PYTHIA}
~\cite{pythia}, and \textsc{CTEQ5L} or \textsc{CTEQ6L}~\cite{cteq} leading-order (LO)
parton distribution functions. The signal cross sections are
normalized to next-to-next-to-leading order (NNLO)
calculations~\cite{nnlo1,nnlo2}, and branching ratios from
\textsc{HDECAY}~\cite{hdecay}. For both CDF and D\O , events from
multijet (instrumental) backgrounds (``QCD production'') are measured
in data with different methods, in orthogonal samples.
For CDF, backgrounds
from other SM processes were generated using \textsc{PYTHIA},
\textsc{ALPGEN}~\cite{alpgen}, \textsc{MC@NLO}~\cite{MC@NLO}
 and \textsc{HERWIG}~\cite{herwig}
programs. For D\O , these backgrounds were generated using
\textsc{PYTHIA}, \textsc{ALPGEN}, and \textsc{COMPHEP}~\cite{comphep},
with \textsc{PYTHIA} providing parton-showering and hadronization for
all the generators.  Background processes were normalized using either
experimental data or next-to-leading order calculations from
\textsc{MCFM}~\cite{mcfm}.

Integrated luminosities, 
and references to the collaborations' public documentation for each analysis
are given in Table~\ref{tab:cdfacc}
for CDF and in Table~\ref{tab:dzacc} for D\O .  
The tables include the ranges of Higgs boson mass ($m_H$) over which
the searches were performed. 

\begin{table}[h]
\caption{\label{tab:cdfacc}Luminosity, explored mass range and references 
for the CDF analyses.  $\ell$ stands for either $e$ or $\mu$.
}
\begin{ruledtabular}
\begin{tabular}{lcccccc}
\\
&$WH\rightarrow \ell\nu b\bar{b}$ & $ZH\rightarrow \nu\bar{\nu} b\bar{b}$ &
$ZH\rightarrow \ell^+\ell^- b\bar{b}$ &  $H\rightarrow W^+ W^- $ &   $H$ + $X
\rightarrow \tau^+ \tau^- $ + 2 jets & 
~~~~~~~~\\ 
                           & 2~(TDT,LDT,STC) &   TDT,LDT               &   ST,DT & 
low,high S/B & H+VBF+WH+ZH &
~~~~~~~~\\ \hline 
Luminosity (\ifb)         & 1.9        &  1.7            & 1.0      & 2.4     & 2.0\\ 
$m_{H}$ range (GeV/c$^2$) & 110-150    & 100-150         & 110-150  & 110-200 & 110-150\\
Reference       & \cite{cdfWH} & \cite{cdfZH}& \cite{cdfZHll} & \cite{cdfHWW} & \cite{cdfHtt} \\
\end{tabular}
\end{ruledtabular}
\end{table}
\vglue 0.5cm 
\begin{table}[h]
\caption{\label{tab:dzacc}Luminosity, explored mass range and references 
for the D\O\ analyses.  $\ell$ stands for either $e$ or $\mu$.
}
\begin{ruledtabular}
\begin{tabular}{lcccccc}
\\
&$WH\rightarrow \ell\nu b\bar{b}$ & $ZH\rightarrow \nu\bar{\nu} b\bar{b}$ &
$ZH\rightarrow \ell^+\ell^- b\bar{b}$ &  $H\rightarrow W^+ W^- $ & 
 $WH \rightarrow WW^+ W^-$ & $H \rightarrow \gamma \gamma$ \\ 
                           &  2~(ST,DT) &   DT               & 2~(ST,DT) & 
$\rightarrow \ell^\pm\nu \ell^\mp\nu$ & 
$\rightarrow \ell^\pm\nu \ell^\pm\nu$  & \\ \hline 
Luminosity (\ifb)         & 1.7        &  2.1            & 1.1 & 2.3 & 1.1 & 2.3\\ 
$m_{H}$ range (GeV/c$^2$) & 105-145    & 105-145         & 105-145  & 110-200 & 120-200 & 105-145\\
Reference       & \cite{dzWHl} & \cite{dzZHv} & \cite{dzZHll} 
& \cite{dzHWW},\cite{dzHWW-2b} 
 & \cite{dzWWW}  & \cite{dzHgg} 
\\
\end{tabular}
\end{ruledtabular}
\end{table}

\section{Combining Channels} 

To gain confidence that the final result does not depend on the
details of the statistical formulation, 
we performed several types of combinations, using the
Bayesian and  Modified Frequentist approaches, which give similar results
(within 10\%).
Both
methods rely on distributions in the final discriminants, and not just on
their single integrated
values.  Systematic uncertainties enter as uncertainties on the
expected number of signal and background events, as well
as on the distribution of the discriminants in 
each analysis (``shape uncertainties'').
Both methods use likelihood calculations based on Poisson
probabilities.

\subsection{Bayesian Method}

Because there is no experimental information on the production cross section for
the Higgs boson, in the Bayesian technique~\cite{CDFhiggs} we assign a flat prior
for the total number of selected Higgs events.  For a given Higgs boson mass, the
combined likelihood is a product of likelihoods for the individual
channels, each of which is a product over histogram bins:

\begin{equation}
{\cal{L}}(R,{\vec{s}},{\vec{b}}|{\vec{n}},{\vec{\theta}})\times\pi({\vec{\theta}})
= \prod_{i=1}^{N_C}\prod_{j=1}^{Nbins} \mu_{ij}^{n_{ij}} e^{-\mu_{ij}}/n_{ij}!
\times\prod_{k=1}^{n_{np}}e^{-\theta_k^2/2}
\end{equation}

\noindent where the first product is over the number of channels
($N_C$), and the second product is over histogram bins containing
$n_{ij}$ events, binned in  ranges of the final discriminants used for
individual analyses, such as the dijet mass, neural-network outputs, 
or matrix-element likelihoods.
 The parameters that contribute to the
expected bin contents are $\mu_{ij} =R \times s_{ij}({\vec{\theta}}) + b_{ij}({\vec{\theta}})$ 
for the
channel $i$ and the histogram bin $j$, where $s_{ij}$ and $b_{ij}$ 
represent the expected background and signal in the bin, and $R$ is a scaling factor
applied to the signal to test the sensitivity level of the experiment.  
Truncated Gaussian priors are used for each of the nuisance parameters                                               
$\theta_k$, which define
the
sensitivity of the predicted signal and background estimates to systematic uncertainties.
These
can take the form of uncertainties on overall rates, as well as the shapes of the distributions
used for combination.   These systematic uncertainties can be far larger
than the expected SM signal, and are therefore important in the calculation of limits. 
The truncation
is applied so that no prediction of any signal or background in any bin is negative.
The posterior density function is
then integrated over all parameters (including correlations) except for $R$,
and a 95\% credibility level upper limit on $R$ is estimated
by calculating the value of $R$ that corresponds to 95\% of the area
of the resulting distribution.

\subsection{Modified Frequentist Method}

The Modified Frequentist technique relies on the $CL_s$ method, using a
log-likelihood ratio (LLR) as test statistic~\cite{DZhiggs}:
\begin{equation}
LLR = -2\ln\frac{p({\mathrm{data}}|H_1)}{p({\mathrm{data}}|H_0)},
\end{equation}
where $H_1$ denotes the test hypothesis, which admits the presence of 
SM backgrounds and a Higgs
boson signal, while $H_0$ is the null hypothesis, for
only SM backgrounds.  The probabilities $p$ are computed using the best-fit
values of the nuisance parameters for each event, separately 
for each of the two hypotheses,
and include the Poisson probabilities of observing the data multiplied by Gaussian
constraints for the values of the nuisance parameters.  This technique
extends the LEP procedure~\cite{pdgstats} which does not involve a fit, in order to
yield better sensitivity when expected signals are small and
systematic uncertainties on backgrounds are large. 

The $CL_s$ technique involves computing two $p$-values, $CL_{s+b}$ and $CL_b$.
The latter is defined by
\begin{equation}
1-CL_b = p(LLR\le LLR_{\mathrm{obs}} | H_0),
\end{equation}
where $LLR_{\mathrm{obs}}$ is the value of the test statistic computed for the
data. $1-CL_b$ is the probability of observing a signal-plus-background-like outcome 
without the presence of signal, i.e. the probability
that an upward fluctuation of the background provides  a signal-plus-background-like
response as observed in data.
The other $p$-value is defined by
\begin{equation}
CL_{s+b} = p(LLR\ge LLR_{\mathrm{obs}} | H_1),
\end{equation}
and this corresponds to the probability of a downward fluctuation of the sum
of signal and background in 
the data.  A small value of $CL_{s+b}$ reflects inconsistency with  $H_1$.
It is also possible to have a downward fluctuation in data even in the absence of
any signal, and a small value of $CL_{s+b}$ is possible even if the expected signal is
so small that it cannot be tested with the experiment.  To minimize the possibility
of  excluding  a signal to which there is insufficient sensitivity 
(an outcome  expected 5\% of the time at the 95\% C.L., for full coverage),
we use the quantity $CL_s=CL_{s+b}/CL_b$.  If $CL_s<0.05$ for a particular choice
of $H_1$, that hypothesis is deemed excluded at the 95\% C.L.

Systematic uncertainties are included  by fluctuating the predictions for
signal and background rates in each bin of each histogram in a correlated way when
generating the pseudoexperiments used to compute $CL_{s+b}$ and $CL_b$.

\subsection{Systematic Uncertainties} 

Systematic uncertainties differ
between experiments and analyses, and they affect the rates and shapes of the predicted
signal and background in correlated ways.  The combined results incorporate
the sensitivity of predictions to  values of nuisance parameters,
and correlations are included, between rates and shapes, between signals and backgrounds,
and between channels within experiments and between experiments.
More on these issues can be found in the
individual analysis notes~\cite{cdfWH}-\cite{dzHgg}.  Here we
consider only the largest contributions and correlations between and
within the two experiments.

\subsubsection{Correlated Systematics between CDF and D\O}
The uncertainty on the measurement of the integrated luminosity is 6\%
(CDF) and 6.1\% (D\O ).  
Of this value, 4\% arises from the uncertainty
on the inelastic \pp~scattering cross section, which is correlated
between CDF and D\O . 
The uncertainty on the production rates for the signal, for 
top-quark processes (\ttbar~and single top) and for electroweak processes
($WW$, $WZ$, and $ZZ$) are taken as correlated between the two
experiments. As the methods of measuring the multijet (``QCD'')
backgrounds differ between CDF and D\O , there is no
correlation assumed between these rates.  Similarly, the large
uncertainties on the background rates for $W$+heavy flavor (HF) and $Z$+heavy flavor
are considered at this time to be uncorrelated, as both CDF and D\O\ estimate
these rates using data control samples, but employ different techniques.
The calibrations of fake leptons, unvetoed $\gamma\rightarrow e^+e^-$ conversions,
$b$-tag efficiencies and mistag rates are performed by each collaboration
using independent data samples and methods, hence are considered uncorrelated.

\subsubsection{Correlated Systematic Uncertainties for CDF}
The dominant systematic uncertainties for the CDF analyses are shown
in Tables~\ref{tab:cdfsystwh1},VI,\ref{tab:cdfllbb1},
\ref{tab:cdfsystww},
\ref{tab:cdfsysttautau}. 
Each source induces a correlated uncertainty across all CDF channels
sensitive to that source.
For \hbb, the largest
uncertainties on signal arise from a scale factor for $b$-tagging 
(5.3-16\%), jet energy scale (1-20\%) and MC modeling (2-10\%). 
The shape dependence of the jet energy scale, $b$-tagging and
uncertainties  on gluon radiation (``ISR'' and ``FSR'')
are taken into account for some analyses
(see tables).
For
\hww, the largest uncertainty comes from
MC modeling (5\%).  For simulated backgrounds, the uncertainties on the
expected rates range from 11-40\% (depending on background). 
The backgrounds with the largest systematic uncertainties                                          
are in general quite small. Such uncertainties are 
constrained by fits to  the nuisance parameters, and they
do not affect the result significantly.
Because
the largest background contributions are measured using data, these
uncertainties are treated as uncorrelated for the \hbb~channels. For
the \hww~channel, the  uncertainty on luminosity is taken to be correlated
between signal and background. The differences in the resulting limits
whether treating the
remaining uncertainties as correlated or uncorrelated, is $5\%$.

\begin{table}[t]
\begin{center}
\caption{Systematic uncertainties on the signal contributions  for CDF's
loose double tag (LDT) channel and tight double-tag (TDT) channel.
Systematic uncertainties are listed by name, see the original references for a detailed explanation of their meaning and on how they are derived. 
Systematic uncertainties for $WH$ shown in this table are obtained for $m_H=115$ GeV/c$^2$.
Uncertainties are
relative, in percent and are symmetric unless otherwise indicated.  }
\vskip 0.5cm                                                             
{\centerline{CDF: Loose Double Tag (LDT)~ $WH$ Analysis}}
\vskip 0.099cm
\label{tab:cdfsystwh1}
\begin{tabular}{|l|c|c|c|c|c|c|} \hline
Contribution   & ~~$W$+HF~~ & ~Mistags~ & ~~~Top~~~ & ~~Diboson~~ & ~~Non-$W$~~ & ~~~~WH~~~~  \\ \hline
Luminosity ($\sigma_{\mathrm{inel}}(p{\bar{p}})$)          & 0      & 0       & 4   & 4       & 0       &    4   \\
Luminosity Monitor        & 0      & 0       & 5   & 5       & 0       &    5   \\
Lepton ID    & 0      & 0       & 2   & 2       & 0       &    2   \\
Jet Energy Scale         & 0      & 0       & 0   & 0       & 0       &    3   \\
Mistag Rate      & 0      &   8    & 0   & 0       & 0       &    0   \\
B-Tag Efficiency      & 0      & 0       & 0   & 0       & 0       &    8   \\
$t{\bar{t}}$ Cross Section         & 0      & 0       & 15 & 0       & 0       &    0   \\
Diboson Rate        & 0      & 0       & 0   & 10      & 0       &    0   \\
NNLO Cross Section           & 0      & 0       & 0   & 0       & 0       &    1 \\
HF Fraction in W+jets          &    42  & 0       & 0   & 0       & 0       &    0   \\
ISR+FSR+PDF      & 0      & 0       & 0   & 0       & 0       &    5 \\ 
QCD Rate         & 0      & 0       & 0   & 0       & 18      &    0   \\
\hline
\end{tabular}
%
\vskip 0.5cm                                                                                                          
{\centerline{CDF: Tight Double Tag (TDT)~ $WH$ Analysis}}
\vskip 0.099cm                                                                          
\begin{tabular}{|l|c|c|c|c|c|c|} \hline
Contribution   & ~~$W$+HF~~ & ~Mistags~ & ~~~Top~~~ & ~~Diboson~~ & ~~Non-$W$~~ & ~~~~WH~~~~  \\ \hline
Luminosity ($\sigma_{\mathrm{inel}}(p{\bar{p}})$)          & 0      & 0       & 4   & 4       & 0       &    4 \\
Luminosity Monitor        & 0      & 0       & 5   & 5       & 0       &    5 \\
Lepton ID    & 0      & 0       & 2   & 2       & 0       &    2 \\
Jet Energy Scale         & 0      & 0       & 0   & 0       & 0       &    3 \\
Mistag Rate      & 0      &  9      & 0   & 0       & 0       &    0 \\
B-Tag Efficiency      & 0      & 0       & 0   & 0       & 0       &   9  \\
$t{\bar{t}}$ Cross Section         & 0      & 0       &  15 & 0       & 0       &    0 \\
Diboson Rate        & 0      & 0       & 0   &   10    & 0       &    0 \\
NNLO Cross Section           & 0      & 0       & 0   & 0       & 0       &    1 \\
HF Fraction in W+jets          &   42   & 0       & 0   & 0       & 0       &    0 \\
ISR+FSR+PDF      & 0      & 0       & 0   & 0       & 0       &  6   \\ 
QCD Rate         & 0      & 0       & 0   & 0       &   18    &    0 \\
\hline
\end{tabular}
\end{center}
\end{table}

\clearpage

\subsubsection{Correlated Systematic Uncertainties for D\O }
The dominant systematic uncertainties for D\O\ analyses are shown in Tables
IV,\ref{tab:d0vvbb},\ref{tab:d0llbb1},\ref{tab:d0systww},\ref{tab:d0systwww},
\ref{tab:d0systgg}.
Each source induces a correlated uncertainty across all D\O\ channels
sensitive to that source.
The \hbb~analyses have an uncertainty on the
$b$-tagging rate of 3-10\% per tagged jet, and  also an
uncertainty on the jet energy and acceptance of 6-9\% (jet
identification or jet ID, energy scale, and jet resolution).
The shape dependence of 
the uncertainty on $W+$ jet modeling 
is taken into account in the limit setting, and has a small effect ($\sim 5\%$) on 
the final result.
For the \hww~and \www, the largest uncertainties are associated with lepton
measurement and acceptance. These values range from 2-11\% depending on
the final state.  The largest contributing factor to all analyses is
the uncertainty on cross sections for simulated background, and is 
6-18\%. 
All systematic uncertainties arising from the
same source are taken to be correlated between the different backgrounds and
between signal and background.
\begin{table}[h]
\begin{center}
\label{tab:d0systwh1}
\caption{Systematic uncertainties on the signal contributions  for D\O 's
$WH\rightarrow\ell\nu b{\bar{b}}$ single (ST) and double tag (DT) channel.
Systematic uncertainties are listed by name, see the original references for a detailed explanation of their meaning and on how they are derived.  
Systematic uncertainties for $WH$ shown in this table are obtained for $m_H=115$ GeV/c$^2$.
  Uncertainties are
relative, in percent and are symmetric unless otherwise indicated.  }
\vskip 0.2cm
{\centerline{D\O : Single Tag (ST) $WH$ Analysis}} 
\vskip 0.099cm
\begin{tabular}{| l |c |c |c |c |c |c |c |}
\hline 
Contribution  &~WZ/WW~&Wbb/Wcc&Wjj/Wcj&$~~~t\bar{t}~~~$&single top&~~QCD~~& ~~~WH~~~\\ 
\hline                                                                            
Luminosity                &  6.1  &  6.1  &  6.1  &  6.1  &  6.1  &  0    &  6.1  \\ 
Trigger eff.              &     3 &     3 &     3 &     3 &     3 &  0    &     3 \\       
Primary Vertex/misc.      &     4 &     4 &     4 &     4 &     4 &  0    &     4 \\       
EM ID/Reco eff./resol.    &     5 &     5 &     5 &     5 &     5 &  0    &     5 \\       
Muon ID/Reco eff./resol.  &     7 &     7 &     7 &     7 &     7 &  0    &     7 \\        
Jet ID/Reco eff.          &     3 &     3 &     3 &     3 &     3 &  0    &     3 \\ 
Jet multiplicity/frag.    &     5 &     5 &     5 &     5 &     5 &  0    &     5 \\       
Jet Energy Scale          &     3 &     4 &     3 &     4 &     2 &  0    &     3 \\       
Jet taggability           &     3 &     3 &     3 &     3 &     3 &  0    &     3 \\ 
NN $b$-tagger Scale Factor&     3 &     3 &  15   &     3 &     3 &  0    &     3 \\ 
Cross Section             &     6 &     9 &     9 &    16 &    16 &  0    &     6 \\       
Heavy-Flavor K-factor     &  0    &    20 &    20 &  0    &  0    &  0    &  0    \\       
Instrumental-WH-1         &  0    &  0    &  0    &  0    &  0    &    19 &  0    \\ 
\hline                                                                            
\end{tabular}                                                                            
\vskip 0.5cm                                                                             
{\centerline{D\O : Double Tag (DT) $WH$ Analysis}}
\vskip 0.099cm
\begin{tabular}{| l |c |c |c |c |c |c |c |}                                                
\hline 
Contribution  &~WZ/WW~&Wbb/Wcc&Wjj/Wcj&$~~~t\bar{t}~~~$&single top&~~QCD~~& ~~~WH~~~\\ 
\hline                                                                            
Luminosity                &  6.1  &  6.1  &  6.1  &  6.1  &  6.1  &  0    &  6.1  \\ 
Trigger eff.              &     3 &     3 &     3 &     3 &     3 &  0    &     3 \\       
Primary Vertex/misc.      &     4 &     4 &     4 &     4 &     4 &  0    &     4 \\       
EM ID/Reco eff./resol.    &     5 &     5 &     5 &     5 &     5 &  0    &     5 \\       
Muon ID/Reco eff./resol.  &     7 &     7 &     7 &     7 &     7 &  0    &     7 \\        
Jet ID/Reco eff.          &     3 &     3 &     3 &     3 &     3 &  0    &     3 \\ 
Jet multiplicity/frag.    &     5 &     5 &     5 &     5 &     5 &  0    &     5 \\       
Jet Energy Scale          &     3 &     4 &     3 &     4 &     2 &  0    &     3 \\       
Jet taggability           &     3 &     3 &     3 &     3 &     3 &  0    &     3 \\ 
NN $b$-tagger Scale Factor&     6 &     6 &  25   &     6 &     6 &  0    &     6 \\ 
Cross Section             &     6 &     9 &     9 &    16 &    16 &  0    &     6 \\       
Heavy-Flavor K-factor     &  0    &    20 &  2    &  0    &  0    &  0    &  0    \\ 
Instrumental-WH-2         &  0    &  0    &  0    &  0    &  0    &     31&  0    \\ 
\hline                                                                            
\end{tabular}                                                                                           
\end{center}                                                                                            \end{table}                                                                                              

\begin{table}
\begin{center}
\caption{Systematic uncertainties on the contributions for D\O 's $ZH\rightarrow \nu \nu b{\bar{b}}$ double-tag (DT) channel.
Systematic uncertainties are listed by name, see the original references for a detailed explanation of their meaning and on how they are derived.  
Systematic uncertainties for $ZH$, $WH$  shown in this table are obtained for $m_H=115$ GeV/c$^2$.
Uncertainties are
relative, in percent and are symmetric unless otherwise indicated. }
\label{tab:d0vvbb}
\vskip 0.5cm                                                                                                          
{\centerline{D\O : Double Tag (DT)~ $ZH \rightarrow \nu\nu b \bar{b}$ Analysis}}
\vskip 0.099cm                                                                                                          
\begin{tabular}{| l | c | c | c | c | c | c |}                                                                               
\hline 
Contribution                       &~WZ/ZZ~&~Z+jets~&~W+jets~ &~~~$t\bar{t}$&~~ZH,WH~~\\ \hline                               
Luminosity                             &  6.1  &  6.1  &  6.1  &  6.1  &  6.1    \\ 
Trigger eff.                           &  5    &  5    &  5    &  5    &  5      \\      
Jet Energy Scale                       &  3    &     3 &     3 &     3 &  2      \\       
Jet ID/resolution.                     &  2    &     2 &     2 &     2 &  2      \\
B-tagging/taggability                  &  6    &  6    &  6    &  6    &  6      \\ 
Cross Section                          &  6    &  15   &  15   & 18    &  6      \\                  
Heavy Flavour K-factor                 &  -    &  50   &  50   &  -    &  -      \\ 
\hline 
\end{tabular}                                                                                                           
\end{center}
\end{table}
\begin{table}
\label{tab:cdfvvbb1}
\caption{Systematic uncertainties for CDF's $ZH\rightarrow\nu{\bar{\nu}} b{\bar{b}}$ loose double tag (LDT) channel and tight double-tag (TDT) channel.
Systematic uncertainties are listed by name, see the original references for a detailed explanation of their meaning and on how they are derived.  
Systematic uncertainties for $ZH$ and $WH$ shown in this table are obtained for $m_H=120$ GeV/c$^2$.
Uncertainties are
relative, in percent and are symmetric unless otherwise indicated.  }
\begin{center}
\vskip 0.5cm                                                                                                          
{\centerline{CDF: Loose Double Tag (LDT)~ $ZH \rightarrow \nu\nu b \bar{b}$ Analysis}}
\vskip 0.099cm    
\begin{tabular}{|l|c|c|c|c|c|c|c|c|c|c|c|}\hline
Contribution & Mistag & ~~QCD~~ &Single top &$~~~~t{\bar{t}}$~~~~& ~~$WW$~~ & ~~$WZ$~~ & ~~$ZZ$~~& $W\rightarrow \ell\nu$ &$Z\rightarrow\ell\ell/\nu\nu$ &$ZH$&$WH$ \\ 
  \hline
Luminosity ($\sigma_{\mathrm{inel}}(p{\bar{p}})$) & 0 & 0 &  4 &    4   &       4   &  4  &  4  &   4  &        4   &   4   &  4   \\           
Luminosity Monitor                                              &  0  &  0  &    5   &    5      &   5   &  5  &  5  &  5  &   5   &     5   &  5       \\               
Trigger (shape dep.)                                            &  0  &  0  &    4   &    3      &   4   &  4  &  4  &  4  &   4   &     4   &  3       \\        
Lepton Veto                                                                &  0  &  0  &    2   &    2      &   2   &  2  &  2  &       2  & 2    &      2   &  2       \\   
Jet Energy Scale (shape dep.)                   &  0  &  0  & $^{+4}_{-7}$ & $^{-11}_{-4}$ & $^{+4}_{-8}$ & $^{+4}_{-8}$ & $^{+9}_{-9}$ & $^{+6}_{-13}$ & $^{+13}_{-9}$ & $^{+3}_{-6}$ & $^{-6}_{-2}$   \\
Mistag Rate ~(shape dep.)                                       &  29 &  0  &    0   &    0      &   0   &  0  & 0      &   0  & 0      &    0  &   0  \\ 
B-Tag Efficiency                                                &  0  &  0  &    9.2 &    9.2    &   9.2 & 9.2 & 9.2  &  9.2  &  9.2   &        9.2     &  9.2  \\
$\sigma(p{\bar{p}}\rightarrow Z+HF)$   &  0  &  0  &    0   &    0      &   0   &  0  &  0  &   0  &   40       &    0  &  0   \\       
$\sigma(p{\bar{p}}\rightarrow W+HF)$   &  0  &  0  &    0   &    0      &   0   &  0  &  0  &   40  &   0       &    0  &  0   \\       
Diboson Cross Section                                           &  0  &  0  &    0   &    0      &  11.5   &  11.5 &  11.5 &   0  &   0 &    0  &  0   \\ 
$t{\bar{t}}$ Cross Section                      &  0  &  0  &    0   &    10     &   0   &  0  &  0  &  0  &   0   &     0   &  0       \\   
Single Top Cross Section                        &  0  &  0  &   11.5   &    0      &   0   &  0  &  0  &        0  &   0   &     0   &  0       \\   
ISR                                                                             & 0   & 0    &  0    &   0       &   0   & 0   & 0      &   0  & 0      &$^{+5.0}_{-1.7}$&$^{+4.9}_{+0.4}$\\
FSR                                                                             & 0   & 0    &  0    &   0       &   0   & 0   & 0      &   0  & 0      &$^{+0.4}_{+4.9}$&$^{+2.5}_{+5.2}$ \\
PDF Uncertainty                                                                 &  0  &  0  &    2   &    2      &   2   &  2  &  2  &  2  &   2   &     2   &  2       \\        
QCD Rate (shape dep.)                                           &  0  &  50  &    0   &    0     &   0   &  0  &  0  &  0  &   0   &     0   &  0       \\ 
\hline          
\end{tabular}
%
\vskip 0.5cm                                                     
{\centerline{CDF: Tight Double Tag (TDT)~ $ZH \rightarrow \nu\nu b \bar{b}$ Analysis}}
\vskip 0.099cm                                                     
\begin{tabular}{|l|c|c|c|c|c|c|c|c|c|c|c|}\hline
Contribution     & Mistag & ~~QCD~~ & Single top  &$ ~~~~t{\bar{t}}$~~~~ &~~$WW$~~& ~~$WZ$~~& ~~$ZZ$~~ & $W\rightarrow \ell\nu$ &$Z\rightarrow\ell\ell/\nu\nu$ &$ZH$ &$WH$ \\ 
\hline
%
Luminosity ($\sigma_{\mathrm{inel}}(p{\bar{p}})$)     &  0  &  0  &       4 &    4  &      4  &  4  &  4  &   4  & 4   &   4   &     4   \\      
Luminosity Monitor                                                                                      &  0  &  0  &       5 &    5  &      5  &  5  &  5  &   5  & 5   &   5   &     5   \\      
Trigger                                                                                                                 &  0  &  0  &       4 &    3  &      5  &  5  &  4  &   4  & 4   &   4   &     4   \\      
Lepton Veto                                                                                                             &  0  &  0  &       2 &    2  &      2  &  2  &  2  &   2  & 2   &   0   &     0      \\   
Jet Energy Scale (shape dep.)                                                                   &  0  &  0  & $^{-5}_{-7}$ & $^{-9}_{-3}$ & $^{+7}_{-8}$ & $^{+7}_{-8}$ & $^{+8}_{-7}$ & $^{+6}_{-12}$ & $^{+10}_{-6}$ & $^{+2}_{-6}$ & $^{+3}_{+7}$    \\       
Mistag Rate-2 (shape dep.)                                                                      & 23  & 0 &   0  &   0       &    0 &  0   &  0   &     0    &     0    &     0     &      0        \\ 
B-Tag Efficiency                                                                                        &  0  &  8.6  &       8.6 &    8.6  &      8.6  &  8.6 &   8.6  &   8.6  & 8.6   &   8.6   &     8.6   \\     
$\sigma(p{\bar{p}}\rightarrow Z+HF)$                                    &  0  &  0  &       0 &    0  &      0  &  0  &  0  &   0  & 40  &   0   &     0   \\      
$\sigma(p{\bar{p}}\rightarrow W+HF)$                                    &  0  &  0  &       0 &    0  &      0  &  0  &  0  &   40  & 0   &   0   &     0   \\      
Diboson Cross Section                                                                                   &  0  &  0  &       0 &    0  &     11.5  &  11.5 &  11.5  &   0  & 0   &   0   &     0   \\      
$t{\bar{t}}$ Cross Section                                                              &  0  &  0  &       0 &    10  &      0  &  0  &  0  &   0  & 0   &   0   &     0   \\      
Single Top Cross Section                                                                                &  0  &  0  &      11.5 &    0  &      0  &  0  &  0  &   0  & 0   &   0   &     0   \\      
ISR                                                                                                                     &  0  &  0  &       0 &    0  &      0  &  0  &  0  &   0  & 0   &$^{+5.0}_{-1.7}$&$^{+4.9}_{+0.4}$\\    
FSR                                                                                                                     &  0  &  0  &       0 &    0  &      0  &  0  &  0  &   0  & 0   &$^{+0.4}_{+4.9}$&$^{+2.5}_{+5.2}$ \\    
PDF                                                                                                                     &  0  &  0  &       2 &    2  &      2  &  2  &  2  &   2  & 2   &   2   &     2   \\      
QCD Rate  (shape dep.)                                                                          &  0  &  50  &       0 &    0  &      0  &  0  &  0  &   0  & 0   &   0   &     0   \\      
\hline  
\end{tabular}
\end{center}
\end{table}
\begin{table}
\begin{center}
\caption{Systematic uncertainties on the contributions for CDF's $ZH\rightarrow \ell^+\ell^-b{\bar{b}}$ single-tag (ST) channel.
Systematic uncertainties are listed by name, see the original references for a detailed explanation of their meaning and on how they are derived.  
Systematic uncertainties for $ZH$  shown in this table are obtained for $m_H=115$ GeV/c$^2$.
Uncertainties are
relative, in percent and are symmetric unless otherwise indicated. }
\label{tab:cdfllbb1}
\vskip 0.8cm                                                                                                          
{\centerline{CDF: Single Tag (ST)~ $ZH \rightarrow \ell\ell b \bar{b}$ Analysis}}
\vskip 0.099cm                                                                                                          
\begin{tabular}{|l|c|c|c|c|c|c|c|c|} \hline
Contribution   & ~Fakes~ & ~~~Top~~~  & ~~$WZ$~~ & ~~$ZZ$~~ & ~$Z+b{\bar{b}}$~ & ~$Z+c{\bar{c}}$~& ~$Z+$mistag~ & ~~~$ZH$~~~ \\ \hline
Luminosity ($\sigma_{\mathrm{inel}}(p{\bar{p}})$)          & 0     &    4 &    4 &    4 &    4           &    4          & 0        &    4  \\
Luminosity Monitor        & 0     &    5 &    5 &    5 &    5           &    5          & 0        &    5  \\
Lepton ID    & 0     &    1 &    1 &    1 &    1           &    1          & 0        &    1  \\
Fake Leptons       & 50    & 0    & 0    & 0    & 0              & 0             & 0        & 0     \\
Jet Energy Scale  (shape dep.)       & 0     & 
  $^{+1.3}_{-2.6}$   & 
  $^{+1.9}_{-4.4}$   & 
  $^{+4.1}_{-4.4}$   & 
  $^{+12.8}_{-12.4}$   & 
  $^{+0.11.3}_{-9.8}$   & 
  0   & 
  $^{+2.3}_{-2.4}$   \\ 
Mistag Rate      & 0     & 0    & 0    & 0    & 0              & 0             &   13     & 0     \\
B-Tag Efficiency      & 0     &    8 &    8 &    8 &    8           &   16          & 0        &    8  \\
$t{\bar{t}}$ Cross Section         & 0     &   20 & 0    & 0    & 0              & 0             & 0        & 0     \\
Diboson Cross Section        & 0     & 0    & 20   & 0    & 0              & 0             & 0        & 0     \\
$\sigma(p{\bar{p}}\rightarrow Z+HF)$      & 0     & 0    & 0    & 0    &  40            & 40           & 0        & 0     \\
ISR (shape dep.)           & 0     & 0    & 0    & 0    & 0              & 0             & 0        &   $^{+1.1}_{+0.4}$     \\
FSR (shape dep.)           & 0     & 0    & 0    & 0    & 0              & 0             & 0        &   $^{-0.7}_{-1.4}$     \\
\hline
\end{tabular}
%
%
%
\vskip 0.8cm                                                                                                          
{\centerline{CDF: Double Tag (DT)~ $ZH \rightarrow \ell\ell b \bar{b}$ Analysis}}
\vskip 0.099cm                                                                                                          
\begin{tabular}{|l|c|c|c|c|c|c|c|c|} \hline
Contribution   & ~Fakes~ & ~~~Top~~~  & ~~$WZ$~~ & ~~$ZZ$~~ & ~$Z+b{\bar{b}}$~ & ~$Z+c{\bar{c}}$~& ~$Z+$mistag~ & ~~~$ZH$~~~ \\ \hline
Luminosity ($\sigma_{\mathrm{inel}}(p{\bar{p}})$)          & 0     &    4 &    4 &    4 &    4           &    4          &    0     &    4  \\
Luminosity Monitor        & 0     &    5 &    5 &    5 &    5           &    5          &    0     &    5  \\
Lepton ID    & 0     &    1 &    1 &    1 &    1           &    1          &    0     &    1  \\
Fake Leptons       & 50    &    0 &    0 &    0 &    0           &    0          &    0     &    0  \\
Jet Energy Scale (shape dep.)         & 0     &   
$^{+0.1}_{-0.1}$    & 
0    & 
$^{+0.5}_{-3.0}$    & 
$^{+3.1}_{-7.8}$    & 
$^{+8.7}_{-0}$        & 
0   & 
  $^{+0.3}_{-1.2}$   \\  
Mistag Rate      & 0     &    0 &    0 &    0 &    0           &    0          &   24     &    0  \\
B-Tag Efficiency      & 0     &   16 &   16 &   16 &   16           &    32         &    0     &   16  \\
$t{\bar{t}}$ Cross Section         & 0     &   20 &    0 &    0 &    0           &    0          &    0     &    0  \\
Diboson Cross Section        & 0     &    0 &   20 &    0 &    0           &    0          &    0     &    0  \\
$\sigma(p{\bar{p}}\rightarrow Z+HF)$      & 0     &    0 &    0 &    0 &   40           &   40          &    0     &    0  \\
ISR (shape dep.)           & 0     & 0    & 0    & 0    & 0              & 0             & 0        &   $^{+4.6}_{+0.6}$     \\
FSR (shape dep.)           & 0     & 0    & 0    & 0    & 0              & 0             & 0        &   $^{+5.3}_{+3.7}$    \\
\hline
\end{tabular}
\end{center}
\end{table}


\begin{table}
\begin{center}
\caption{Systematic uncertainties on the contributions for D\O 's $ZH\rightarrow \ell^+\ell^-b{\bar{b}}$ single-tag (ST) channel.
Systematic uncertainties are listed by name, see the original references for a detailed explanation of their meaning and on how they are derived.  
Systematic uncertainties for $ZH$  shown in this table are obtained for $m_H=115$ GeV/c$^2$.
Uncertainties are relative, in percent and are symmetric unless otherwise indicated. }
\label{tab:d0llbb1}
\vskip 0.8cm                                                                                                          
{\centerline{D\O : Single Tag (ST)~ $ZH \rightarrow \ell\ell b \bar{b}$ Analysis}}
\vskip 0.099cm                                                                                                          
\begin{tabular}{ | l | c | c | c | c | c | c | c |}                                                                               
\hline      
Contribution & ~~WZ/ZZ~~ &~~Zbb/Zcc~~&~~~Zjj~~~ &~~~~$t\bar{t}$~~~~&   ~~QCD~~ & ~~~ZH~~~\\ \hline                               
Luminosity                             &  6.1  &  6.1  &  6.1  &  6.1  &  0    &  6.1  \\ 
EM ID/Reco eff.                        &  4    &  4    &  4    &  4    &  0    &  4    \\                                      
Muon ID/Reco eff.                      &  4    &  4    &  4    &  4    &  0    &  4    \\                                      
Jet ID/Reco eff.                       &  2    &  1.5  &  2    &  1.5  &  0    &  1.5  \\ 
Jet Energy Scale (shape dep.)     &  4    &  8    & 11    &  2    &  0    &  2    \\                                      
B-tagging/taggability                  &  7    &  6    &  9    &  3    &  0    &  3    \\ 
Cross Section                          &  7    &  0    &  0    & 18    &  0    &  6    \\                                      
Heavy-Flavor K-factor                  &  0    & 30    & 15    &  0    &  0    &  0    \\ 
Instrumental-ZH-1                      &  0    &  0    &  0    &  0    & 50    &  0    \\ \hline               
\end{tabular}                                                                                                           
\vskip 0.8cm                                                                                                          
{\centerline{D\O : Double Tag (DT)~ $ZH \rightarrow \ell\ell b \bar{b}$ Analysis}}
\vskip 0.099cm                                                                                                          
\begin{tabular}{ | l | c | c | c | c | c | c | c |}                                                                               
\hline 
Contribution & ~~WZ/ZZ~~ &~~Zbb/Zcc~~&~~~Zjj~~~ &~~~~$t\bar{t}$~~~~&   ~~QCD~~ & ~~~ZH~\\ \hline                               
Luminosity                             &  6.1  &  6.1  &  6.1  &  6.1  &  0    &  6.1  \\ 
EM ID/Reco eff.                        &  4    &  4    &  4    &  4    &  0    &  4    \\                                      
Muon ID/Reco eff.                      &  4    &  4    &  4    &  4    &  0    &  4    \\                                      
Jet ID/Reco eff.                       &  2    &  1.5  &  2    &  1.5  &  0    &  1.5  \\ 
Jet Energy Scale (shape dep.)     &  4    &  8    & 11    &  2    &  0    &  2    \\                                      
B-tagging/taggability                  &  8    &  8    &  9    &  7    &  0    &  7    \\ 
Cross Section                          &  7    &  0    &  0    & 18    &  0    &  6    \\                                      
Heavy-Flavor K-factor                  &  0    & 30    & 15    &  0    &  0    &  0    \\ 
Instrumental-ZH-2                      &  0    &  0    &  0    &  0    & 50    &  0    \\ \hline               
\end{tabular}                                                                                                           
\end{center}                                                                                                            
\end{table}

\clearpage

\begin{table}
\begin{center}
\caption{Systematic uncertainties on the contributions for CDF's
$H\rightarrow W^+W^-\rightarrow\ell^{\pm}\ell^{\prime \mp}$ channel.
Systematic uncertainties are listed by name, see the original references for a detailed explanation of their meaning and on how they are derived.  
Systematic uncertainties for $H$ shown in this table are obtained for $m_H=160$ GeV/c$^2$.
Uncertainties are relative, in percent and are symmetric unless otherwise indicated.   The systematic uncertainty 
called ``Normalization''  includes effects of the inelastic $p{\bar{p}}$ cross section, the luminosity monitor acceptance, and the lepton trigger
acceptance. It is considered to be entirely correlated with the luminosity uncertainty.
}
\label{tab:cdfsystww}
\vskip 0.8cm                                                                                                          
{\centerline{CDF:   $H \rightarrow WW \rightarrow \ell^{\pm} \ell^{\prime \mp}$ Analysis}}
\vskip 0.099cm                                                                                                          
\begin{tabular}{|l|c|c|c|c|c|c|c|c|c|} \hline
Contribution    &~~~$WW$~~~& ~~~$WZ$~~~  & ~~~$ZZ$~~~ & ~~~~$t\bar{t}$~~~~ & ~~~~DY~~~~    & ~~$W\gamma$ & $W+$jets~~ &   ~~~~~$H$~~~~~  \\ \hline
Trigger                           &  2    &  2    &   2  &    2  &    3  &    7     &    --     &    3    \\
Lepton ID            .            &  2    &  1    &  1   &    2  &    2  &    1     &    --     &    2    \\
Acceptance                        &    6  &   10  &   10 &   10  &    6  &   10     &    --     &   10    \\
\MET Modeling                     &    1  &    1  &    1 &    1  &    20 &   1      &    --     &    1    \\
Conversions                       &    0  &    0  &    0 &    0  &    0  &    20    &    --     &    0    \\
NNLO Cross Section                &   10  &   10  &   10 &   15  &    5  &    10    &    --     &   10    \\
PDF Uncertainty                   &    2  &    3  &    3 &    2  &    4  &    2     &    --     &    2    \\
Normalization~~~~                 &    6  &    6  &    6 &    6  &    6  &    6     &   23      &    6    \\ \hline
\end{tabular}
\end{center}
\end{table}

\begin{table}
\begin{center}
\caption{Systematic uncertainties on the contributions for D\O 's
$H\rightarrow WW \rightarrow\ell^{\pm}\ell^{\prime \mp}$ channel.
Systematic uncertainties are listed by name, see the original references for a detailed explanation of their meaning and on how they are derived. 
Systematic uncertainties shown in this table are obtained for the $m_H=160$ GeV/c$^2$ Higgs selection.
Uncertainties are relative, in percent and are symmetric unless otherwise indicated.   }
\label{tab:d0systww}
\vskip 0.8cm                                                                                                          
{\centerline{D\O : $H\rightarrow WW \rightarrow\ell^{\pm}\ell^{\prime \mp}$ Analysis }}
\vskip 0.099cm      
\begin{tabular}{| l | c | c | c | c | c | c | c|}                                                                               
\hline
Contribution & Diboson & ~~$Z/\gamma^* \rightarrow \ell\ell$~~&$~~W+jet/\gamma$~~ &~~~~$t\bar{t}~~~~$    & ~~~~QCD~~~~  & ~~~~$H$~~~~      \\ 
\hline                                                                                                                                   
Trigger                          &  5           &   5           & 5             & 5            & --   &   5            \\ 
Lepton ID            .           &  8--13       &  8--13        & 8--13         & 8--13        & --   &   8--13        \\ 
Momentum resolution              &  2--11       &   2--11       & 2--11         & 2--11        & --   &   2--11        \\ 
Jet Energy Scale                 &  1           &   1           & 1             & 1            & --   &   1            \\ 
Cross Section                    &  7           &   6           & 6             & 18           & --   &   10            \\ 
PDF Uncertainty~~~~~             &  4           &   4           & 4             & 4            & --   &   4            \\ 
Normalization~~~~~               &  6           &   6           & 20            & 6            & 30   &   --           \\ 
\hline                                                                                                                                   
\end{tabular}                                                                                                           
\end{center}
\end{table}

\begin{table}
\begin{center}
\caption{Systematic uncertainties on the contributions for CDF's
$ H\rightarrow \tau^+\tau^-$ channels.
Systematic uncertainties are listed by name, see the original references for a detailed explanation of 
their meaning and on how they are derived.  %
Uncertainties with provided shape systematics are labeled with ``s''.
Systematic uncertainties for $H$ shown in this table are obtained for $m_H=115$ GeV/c$^2$.
Uncertainties are relative, in percent and are symmetric unless otherwise indicated.   The systematic uncertainty 
called ``Normalization''  includes effects of the inelastic $p{\bar{p}}$ cross section, the luminosity monitor acceptance, and the lepton trigger
acceptance. It is considered to be entirely correlated with the luminosity uncertainty.
}
\label{tab:cdfsysttautau}
\vskip 0.8cm                                                                                                          
{\centerline{CDF:   $H \rightarrow \tau^+\tau^-$ Analysis}}
\vskip 0.099cm                                                                                                          
\begin{tabular}{|l|c|c|c|c|c|c|c|c|c|c|} \hline
Contribution & ~~$Z/\gamma^* \rightarrow \tau\tau$~~ & ~~$Z/\gamma^* \rightarrow \ell\ell$~~&~~~$t\bar{t}~~~$&diboson  & ~jet $\rightarrow \tau$ 
& W+jet & ~~~$WH$~~~      & ~~~$ZH$~~~      & ~~~VBF~~~      & ~~~$H$~~~    \\  
\hline
Luminosity~~~~                &  6   &  6   &  6   &  6  &  -   &    -     &  6  &  6  &  6  &  6   \\ 
$e,\mu$ Trigger               &  1   &  1   &  1   &  1  &  -   &    -     &  1  &  1  &  1  &  1   \\
$\tau$  Trigger               &  3   &  3   &  3   &  3  &  -   &    -     &  3  &  3  &  3  &  3   \\
$e,\mu,\tau$ ID      .        &  3   &  3   &  3   &  3  &  -   &    -     &  3  &  3  &  3  &  3   \\
PDF Uncertainty               &  1   &  1   &  1   &  1  &  -   &    -     &  1  &  1  &  1  &  1   \\
ISR/FSR                       &  -   &  -   &  -   &  -  &  -   &    -     & 2/0 & 1/1 & 3/1 & 12/1 \\
JES (shape)                   & 16   & 13   &  2   & 10  &  -   &    -     &  3  &  3  &  4  & 14   \\
Cross Section or Norm.         &  2  &  2   & 13   & 10  &  -   &   15     &  5  &  5  & 10  & 10   \\
MC model                      & 20   & 10   &  -   &  -  &  -   &    -     &  -  &  -  &  -  &  -   \\
\hline
\end{tabular}
\end{center}
\end{table}

\begin{table}
\begin{center}
\caption{Systematic uncertainties on the contributions for D\O 's
$WH \rightarrow WWW \rightarrow\ell^{\prime \pm}\ell^{\prime \pm}$ channel.
Systematic uncertainties are listed by name, see the original references for a detailed explanation of their meaning and on how they are derived. 
Systematic uncertainties for $WH$ shown in this table are obtained for $m_H=160$ GeV/c$^2$.
Uncertainties are relative, in percent and are symmetric unless otherwise indicated.   }
\label{tab:d0systwww}
\vskip 0.8cm                                                                                                          
{\centerline{D\O : $WH \rightarrow WWW \rightarrow\ell^{\pm}\ell^{\prime\pm}$ Analysis.}}
\vskip 0.099cm                                                                                                          
\begin{tabular}{| l | c | c | c | c | }
\hline
Contribution                           & ~~WZ/ZZ~~ & Charge flips & ~~~QCD~~~ &~~~~~WH~~~~~  \\ \hline
Trigger eff.                           &  5    &  0                     &  0  &  5    \\
Lepton ID/Reco. eff                    & 10    &  0                     &  0  & 10    \\
Cross Section                          &  7    &  0                     &  0  &  6    \\
Normalization                          &  6    &  0                     &  0  &  0    \\ 
Instrumental-ee ($ee$ final state)                                                    
                                       &  0    &  32                    &  15 &  0    \\
Instrumental-em ($e\mu$ final state)                                                  
                                       &  0    &  0                     &  18 &  0    \\
Instrumental-mm ($\mu\mu$ final state)                                                 
                                       &  0    &  $^{+290}_{-100}$      & 32  &  0    \\ \hline
\end{tabular}
\end{center}
\end{table}

\begin{table}
\begin{center}
\caption{Systematic uncertainties on the contributions for D\O 's
$ H\rightarrow \gamma \gamma$ channels.
Systematic uncertainties are listed by name, see the original references for a detailed explanation of 
their meaning and on how they are derived.  %
Uncertainties are relative, in percent and are symmetric unless otherwise indicated.   
}
\label{tab:d0systgg}
\vskip 0.8cm                                                                                                          
{\centerline{D\O :   $H \rightarrow \gamma \gamma$ Analysis}}
\vskip 0.099cm                                                                                                          
\begin{tabular}{|l|c|c|} \hline
Contribution &  background  & ~~~$H$~~~    \\  
\hline
Luminosity~~~~                            &  6    &  6    \\ 
ID efficiency                             &  1    &  1    \\
Acceptance           .                    &  -    &  2    \\
$\gamma$-jet and jet-jet fakes            &  26   &  -    \\
electron track-match inefficiency         & 10--15& -     \\
Cross Section ($Z$)                       &  4    &  6    \\
Cross Section (QCD $\gamma \gamma$)       &  20   &  -    \\
\hline
\end{tabular}
\end{center}
\end{table}

\clearpage

 \begin{figure}[t]
 \begin{centering}
 \includegraphics[width=14.0cm]{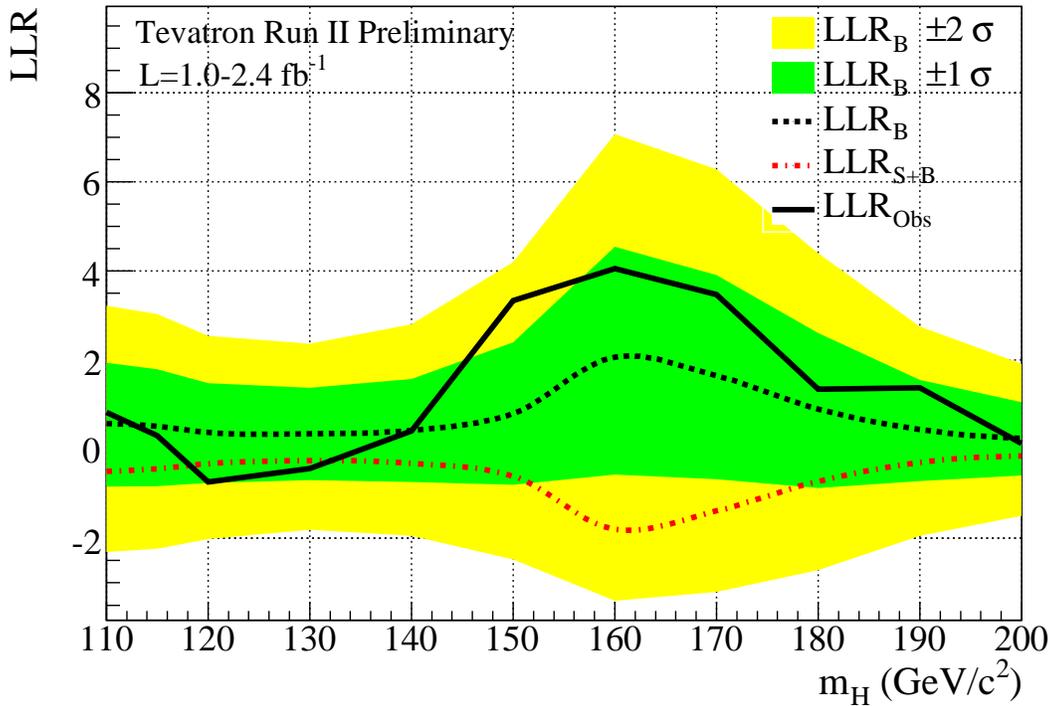}
 \caption{
 \label{fig:comboLLR}
 Distributions of LLR as a function of Higgs mass 
 for the combination of all
 CDF and D\O\ analyses. }
 \end{centering}
 \end{figure}

\vspace*{1cm}
\section{Combined Results} 

Before extracting the combined limits we study the distributions 
of the 
log-likelihood ratio (LLR) for different hypothesis,
to check the expected
sensitivity across the mass range tested.
Figure~\ref{fig:comboLLR}
displays the LLR distributions
for the combined
analyses as a function of $m_{H}$. Included are the results for the
background-only hypothesis (LLR$_{b}$), the signal and background
hypothesis (LLR$_{s+b}$), and for the data (LLR$_{obs}$).  The
shaded bands represent the 1 and 2 standard deviation ($\sigma$)
departures for LLR$_{b}$. 

These
distributions can be interpreted as follows:
The separation between LLR$_{b}$ and LLR$_{s+b}$ provides a
measure of the discriminating power of the search; 
 the size of the 1- and 2-$\sigma$ LLR$_{b}$ bands
provides an estimate of how sensitive the
analysis is to a signal-plus-background-like fluctuation in data, taking account of
the systematic uncertainties;
the value of LLR$_{obs}$ relative to LLR$_{s+b}$ and LLR$_{b}$
indicates whether the data distribution appears to be more signal-plus-background-like
(i.e. closer to the LLR$_{s+b}$ distribution, which is negative by
construction)
or background-like; the significance of any departures
of LLR$_{obs}$ from LLR$_{b}$ can be evaluated by the width of the
LLR$_{b}$ bands.

Using the combination procedures outlined in Section III, we extract limits on
SM Higgs boson production $\sigma \times B(H\rightarrow X)$ in
\pp~collisions at $\sqrt{s}=1.96$~TeV. 
To facilitate comparisons with the standard model and to accommodate analyses with
different degrees of sensitivity, we present our results in terms of
the ratio of obtained limits  to  cross section in the SM, as a function of
Higgs boson mass, for test masses for which
both experiments have performed dedicated searches in different channels.
A value of the combined limit ratio which is less than one would indicate that
that particular Higgs boson mass is excluded at the
 A value  $<1$ would indicate a Higgs boson mass excluded at
95\% C.L. The expected and observed 95\% C.L. ratios to the
SM cross section for the combined CDF and D\O\ analyses are shown in
Figure~\ref{fig:comboRatio}.  The observed and median expected limit ratios
are listed for the tested Higgs boson masses in Table~\ref{tab:ratios}, with
observed (expected) values of 
3.7 (3.3) at $m_{H}=115$~GeV/c$^2$ and
1.1 (1.6) at $m_{H}=160$~GeV/c$^2$.

These results represent about a 40\% improvement in expected sensitivity over
those obtained on the combinations of results of each single experiment,
which yield  observed (expected) limits on the SM  ratios of 
5.0~(4.5) for CDF and 6.4~(5.5) for D\O\ at $m_{H}=115$~GeV/c$^2$, and of 
1.6~(2.6) for CDF and 2.2~(2.4) for D\O\ at $m_{H}=160$~GeV/c$^2$.

\begin{table}[ht]
\caption{\label{tab:ratios} Median expected and observed 95\% CL
cross section ratios for the combined CDF and D\O\ analyses as a function
the Higgs boson mass in GeV/c$^2$.}
\begin{ruledtabular}
\begin{tabular}{lccccccccccc}
%
%
                 & 110 & 115  & 120 & 130 & 140 & 150 & 160 & 170 & 180 & 190 & 200\\ \hline 
Expected         & 3.1 &  3.3 & 3.8 & 4.2 & 3.5 & 2.7 & 1.6 & 1.8 & 2.5 & 3.8 & 5.1\\
Observed         & 2.8 &  3.7 & 6.6 & 5.7 & 3.5 & 2.3 & 1.1 & 1.3 & 2.4 & 2.8 & 5.2\\
%
%
%
\end{tabular}
\end{ruledtabular}
\end{table}
\begin{figure}[ht]
\begin{centering}
\includegraphics[width=14.5cm]{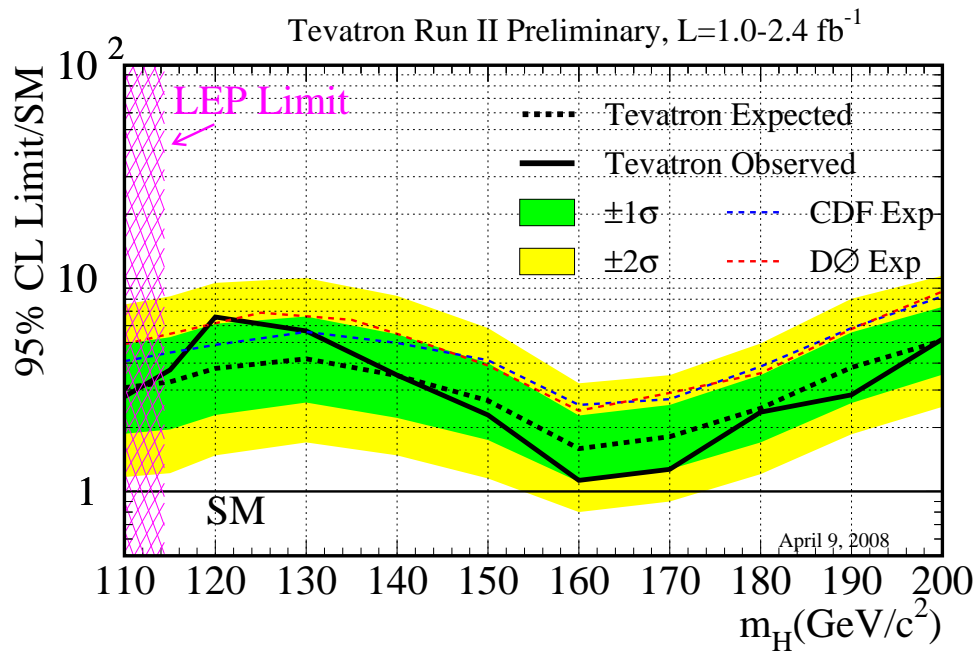}
\caption{
\label{fig:comboRatio}
Observed and expected (median, for the background-only hypothesis)
  95\% C.L. upper limits on the ratios to the
SM cross section, 
as functions of the Higgs boson test mass, 
for the combined CDF and D\O\ analyses.
The limits are expressed as a multiple of the SM prediction
for test masses for which both experiments have performed dedicated
searches in different channels.
The $WH/ZH$ with $H \to b\bar{b}$ and the $\tau\tau$ / $\gamma\gamma$
channels are contributing for $m_H \le 150$ GeV. 
The $H \to WW$ and $WH \to WWW$ channels are contributing
for $m_H \ge 115$ GeV.
The points are joined by straight lines 
for better readability.
  The bands indicate the
68\% and 95\% probability regions where the limits can
fluctuate, in the absence of signal.
Also shown are the  expected upper limits obtained for  
all combined CDF channels, and for  all combined D\O\ channels.
}
\end{centering}
\end{figure}




\clearpage
\begin{thebibliography}{000}

\bibitem{CDFhiggs} CDF Collaboration, ``Combined Upper Limit on
Standard Model Higgs Boson Production'',
CDF Conference Note 8941.

\bibitem{DZhiggs}
\DZero Collaboration, 
``Combined upper limits on standard model Higgs boson production from the D0 experiment with 1.1-2.4 fb$^{-1}$'' 
D\O\ Conference Note 5625.

\bibitem{tev-comb-2007}
CDF and \DZero Collaborations,
``Combined CDF and \DZ Upper Limits on Standard Model Higgs-Boson Production'',
{FERMILAB-PUB-07-656-E}, arXiv:0712.2383

\bibitem{talk-moriond-qcd}
R. Mommsen for the CDF and \DZero Collaborations,
XLIII$^{th}$ Rencontres de Moriond on QCD and High Energy Interactions, 2008,
http://moriond.in2p3.fr/QCD/2008/TuesdayAfternoon/Mommsen.pdf


\bibitem{cdfWH} CDF Collaboration, "Search for Higgs Boson Production
in Association with W Boson with 1.7~\ifb", 
CDF Conference Note 8957.

\bibitem{cdfZH}  CDF Collaboration, 
``Search for the Standard Model Higgs Boson in the Missing Et and B-jets Signature'',
CDF Conference Note 8973.

\bibitem{cdfZHll} CDF Collaboration, 
``Search for ZH in 1 fb-1'',
CDF Conference Note 8742.

\bibitem{cdfHWW} CDF Collaboration, ``Search for $H \rightarrow WW$ Production Using 1.9~fb$^{-1}$'',
CDF Conference Note 8923.

\bibitem{cdfHtt} CDF Collaboration, ``Search for SM Higgs using tau leptons using 2~fb$^{-1}$'',
CDF Conference Note 9179.

\bibitem{dzWHl} D\O\ Collaboration, ``Search for WH Production at
$\sqrt{s}=1.96$~TeV with Neural Networks,'' D\O\ Conference Note 5472.

\bibitem{dzZHv} D\O\ Collaboration, ``Search for the standard model Higgs boson in the $HZ \rightarrow b \bar{b} \nu 
\nu$ channel in 2.1 fb${-1}$ of ppbar collisions at 
$\sqrt{s}=1.96$~TeV'', D\O\  Conference note 5586.

\bibitem{dzZHll} D\O\ Collaboration, ``A Search for \ZHll\  Production at
  D\O\ in \pp\ Collisions at $\sqrt{s}=1.96$~TeV'',
D\O\ Conference Note 5482.


%



\bibitem{dzHWW} D\O\ Collaboration, ``Search for the Higgs boson in
$H \rightarrow W W^* \rightarrow l^+ l^- (\ell,\ell^\prime= e \mu)$ decays with
1.7 ~\ifb~at D\O\ in Run II'', D\O\ Conference Note 5537.

\bibitem{dzHWW-2b} D\O\ Collaboration, ``Search for the Higgs boson in
$H \rightarrow W W^*$  decays with 1.2~\ifb~at D\O\
in Run IIb'', D\O\ Conference Note 5624.


\bibitem{dzWWW} D\O\ Collaboration, ``Search for associated Higgs boson
production $WH\rightarrow WWW^* \rightarrow \ell^\pm \nu
\ell^{\prime\pm} \nu^\prime +X$ in $p\bar{p}$ collisions at
$\sqrt{s}=1.96$~TeV'', 
D\O\ Conference Note 5485.

\bibitem{dzHgg} D\O\ Collaboration, ``Search for a light Higgs boson
in   $\gamma \gamma$ final state'', 
D\O\ Conference Note 5601.

\bibitem{pythia} 
T.~Sjostrand, L.~Lonnblad and S.~Mrenna,
   ``PYTHIA 6.2: Physics and manual,''
  arXiv:hep-ph/0108264.

\bibitem{cteq} 
H.~L.~Lai {\it et al.}, ``Improved Parton
Distributions from Global Analysis of Recent Deep Inelastic Scattering
and Inclusive Jet Data'', Phys. Rev D \textbf{55}, 1280 (1997).

\bibitem{nnlo1} 
S.~Catani, D.~de Florian, M.~Grazzini and P.~Nason,
   ``Soft-gluon resummation for Higgs boson production at hadron colliders,''
  JHEP {\bf 0307}, 028 (2003)
  [arXiv:hep-ph/0306211].

\bibitem{nnlo2} 
K.~A.~Assamagan {\it et al.}  [Higgs Working Group Collaboration],
   ``The Higgs working group: Summary report 2003,''
  arXiv:hep-ph/0406152.

\bibitem{hdecay}
A.~Djouadi, J.~Kalinowski and M.~Spira,
   ``HDECAY: A program for Higgs boson decays in the standard model and its
   supersymmetric extension,''
  Comput.\ Phys.\ Commun.\  {\bf 108}, 56 (1998)
  [arXiv:hep-ph/9704448].

\bibitem{alpgen}
M.~L.~Mangano, M.~Moretti, F.~Piccinini, R.~Pittau and A.~D.~Polosa,
   ``ALPGEN, a generator for hard multiparton processes in hadronic
   collisions,''
  JHEP {\bf 0307}, 001 (2003)
  [arXiv:hep-ph/0206293].

\bibitem{MC@NLO}
S. Frixione and B.R. Webber, 
JHEP 06, 029 (2002)   [arXiv:hep-ph/0204244]

\bibitem{herwig} 
G.~Corcella {\it et al.},
   ``HERWIG 6: An event generator for hadron emission reactions with
   interfering gluons (including supersymmetric processes),''
  JHEP {\bf 0101}, 010 (2001)
  [arXiv:hep-ph/0011363].

\bibitem{comphep}
A.~Pukhov {\it et al.},
   ``CompHEP: A package for evaluation of Feynman diagrams and integration  over
   multi-particle phase space. User's manual for version 33,''
  [arXiv:hep-ph/9908288].

\bibitem{mcfm} J.~Campbell and R.~K.~Ellis, 
 http://mcfm.fnal.gov/. 
%


\bibitem{pdgstats}
T. Junk, Nucl. Instrum. Meth. A434, p. 435-443, 1999,
A.L.~Read, "Modified frequentist analysis of search results (the $CL_s$ method)", in                              
F.~James, L.~Lyons and Y.~Perrin (eds.), {\sl Workshop on Confidence Limits},                                   
CERN, Yellow Report 2000-005, available through {\tt cdsweb.cern.ch}.





\end{thebibliography}
\end{document}

Limits observed and expected for CDF search channels (multiples of SM)

HWW

mh    obs    exp
110  151.2  122.6
120  33.9   37.4
130  17.0  17.4
140  9.5   10.7
150  5.7  8.0
160  3.4  4.8
170  3.3  4.9
180  6.8  6.6
190  14.6  9.8
200  18.4  12.9

bb+MET   (WH and ZH signal together)

mh    obs   exp
110   18.5  9.3
115   19.7  9.7
120   22.6  11.5
125   26.6  13.4
130   33.4  16.6
135   43.0  21.0
140   61.5  31.5
150   127.0  72.1

WH->lvbb

mh   obs   exp
110  8.8   8.6
115  10.1  10.0
120  12.0  11.9
130  18.8  17.4
140  40.0  33.0
150  113.0  80.6

ZH->llbb

mh   obs   exp
100  12    13
110  14    14
115  16    16
120  18    18
130  29    28
140  65    54
150  160   140

       105   115     125   135     145
Obs (11.2)  (17.8)  (30.4) (51.4) (103.5)
Exp (14.9)  (20.4)  (27.3) (42.8)  (88.0)

exp  2.8     2.5     2.3   2.0
     2.6     2.7     2.5   2.3                                                                                                            
HWW_v1.7.pdf
mh [GeV]    120  140 160 180 200
exp
combination 28.7 8.3 3.5 5.3 11.7
obs
combination 48.9 12.3 3.1 5.5 11.4

\ $m_H$   \      &  \ expected \ & \ observed \ \\
\ (GeV) \  & \ limit (pb) \ & \ limit (pb)  \  \\ \hline
105   &  1.29   &   1.42 \\
115   &  1.16   &   1.42 \\
125   &  1.12   &   1.41 \\
135   &  0.94   &   1.16 \\
145   &  0.84   &   1.06 \\ \hline

\vglue 0.2cm 
\begin{table}[h]
\caption{\label{tab:dzacc1}The luminosity, mass range explored and reference 
for the D\O\ \hbb~analyses.   $\ell$ stands for $e$ or $\mu$.
}
\begin{ruledtabular}
\begin{tabular}{lcccccccc}
\\
&$WH\rightarrow e\nu b\bar{b}$ & $WH\rightarrow \mu\nu b\bar{b}$  & \lmet  & $ZH\rightarrow \nu\bar{\nu} b\bar{b}$ & $ZH\rightarrow \ell^+\ell^- b\bar{b}$ \\ 
& DT(ST) & DT(ST) &DT(ST) &  DT(ST) & \\\hline
Luminosity (\ifb)         & 1.7 & 1.7 & 0.9 & 0.9& 1.1\\ 
$m_{H}$ range (GeV/c$^2$)       & 105-145 & 105-145 & 105-135 &105-135  & 105-145\\
Reference       & \cite{dzWHl} & \cite{dzWHl}& \cite{dzZHv} & \cite{dzZHv} & \cite{dzZHll} \\
\end{tabular}
\end{ruledtabular}
\end{table}
\vglue 0.2cm 
\begin{table}[h]
 \caption{\label{tab:dzacc3}The luminosity, mass range explored, and reference 
for the D\O\ \www\ and \hww~analyses.  
}
\begin{ruledtabular}
\begin{tabular}{lccc}
\hline
 &$WW^+ W^- \rightarrow e^\pm\nu e^\pm\nu$
&$H\rightarrow W^+ W^- \rightarrow e^+\nu e^-\nu$ 
&$H\rightarrow W^+ W^- \rightarrow e^\pm\nu \mu^\mp\nu$ \\
 &$WW^+ W^- \rightarrow  e^\pm\nu \mu^\pm\nu$
&$H\rightarrow W^+ W^- \rightarrow  e^\pm\nu \mu^\mp\nu$ 
& \\ 
 &$WW^+ W^- \rightarrow \mu^\pm\nu \mu^\pm\nu$
&$H\rightarrow W^+ W^- \rightarrow \mu^+\nu \mu^-\nu$ 
& \\  
\hline 
Luminosity (\ifb)         & 1.1 &  1.0 & 0.6\\ 
$m_{H}$ range (GeV/c$^2$)       & 120-200 & 120-200 & 120-200\\
Reference       & \cite{dzWWW} & \cite{dzHWWee,dzHWWmm} & \cite{dzHWWem-2b} \\
\end{tabular}
\end{ruledtabular}
\end{table}